\documentclass{aa}
\usepackage{graphicx}
\usepackage{natbib}
\usepackage{color}
\bibpunct{(}{)}{;}{a}{}{,} 
\begin{document}

\def\ms{M$_{\odot}$}
\def\zs{Z$_{\odot}$}

\title{The metallicity distribution of the halo and the satellites of
the Milky way in the hierarchical merging paradigm }

\author{ Nikos Prantzos\inst{1}}
        
\authorrunning{N. Prantzos}
 
\titlerunning{HMD}


\institute{ Institut d'Astrophysique de Paris, UMR7095 CNRS, 
Univ.P. \& M.Curie,  98bis Bd. Arago, 75104 Paris, France, 
                \email{prantzos@iap.fr}
           }
\date{}

\abstract{}{To account for the observed differential metallicity distribution (DMD) of the Milky Way halo, a semi-analytical model is presented in the framework of the hierarchical merging paradigm for structure formation.}
{It is assumed that the Milky Way halo is composed of a number of sub-haloes with properties either as observed in the dwarf satellite galaxies of the Local group (shape of metallicity distribution, effective yield) or derived from calculations of structure formation (sub-halo distribution function).}
{With reasonable assumptions for the parameters involved, we find that the overall shape and effective yield of the Galactic halo DMD can be reproduced in the framework of such a simple model. The low metallicity tail of  the DMD presents a defficiency of stars with respect to the simple model predictions (akin to the G-dwarf problem in the solar neighborhood); it is suggested that an early infall phase can account for that problem, as well as for the observed DMDs of dwarf satellite galaxies.}
{Accretion of galaxies similar (but not identical) to the progenitors of present day dwarf satellites of the Milky Way may well have formed the Galactic halo.
}


\maketitle

\section{Introduction}
The metallicity distribution 
of long-lived stars is one of the most powerful probes of galactic chemical evolution. Indeed, the shape of the metallicity distribution is independent of the rate at which stars are formed. It depends essentially on the stellar initial mass function (IMF), which fixes the ``intrinsic'' (true) yield of the stellar population and the loss or gain of (gaseous) mass from the galaxian system (outflow or inflow, see next section and Prantzos 2007a). In the case of the solar neighborhood, for instance, the shape of the metallicity distribution suggests that the local disk was constructed by infall on long timescales (e.g. Boissier and Prantzos 1999).

The regular shape of the metallicity distribution of the Milky Way (MW) halo can readily be explained by the simple model of galactic chemical evolution with outflow, as suggested by Hartwick (1976, see also Sec. 2.1). However, that explanation lies within the framework of the monolithic collapse scenario for the formation of the MW (Eggen, Lynden-Bell and Sandage, 1962). It seems dificult to keep that interpretation in the currently dominant paradigm of hierarchical galaxy formation. A coherent interpretation should be sought in the framework of the new paradigm, according to which the MW halo was formed from the hierarchical merging of smaller components (sub-haloes).

Several attempts to account for the metallicity distribution of the MW halo  
in the
modern framework were undertaken in recent years, starting with the  work of Bekki and Chiba (2001). They model a galaxy closely ressembling the MW and  despite the low spatial resolution of their simulations (2 kpc), they are able to provide a halo metallicity distribution with a peak at the observed metallicity ([Fe/H]=--1.6), but with a shape which is much narrower than the observed one ; theyobtain a double peaked metallicity distribution (which is not observed), probably reflecting late major mergers due to the small mass resolution. In their five MW-type simulations (in a cosmological framework and using various prescriptions for ouflow/feedback) Scanapieco and Broadhurst (2001) find no ressemblance of the resulting halo metallicity distributions with the observed one. This is also the case in Font et al. (2006), who find systematically too metal rich haloes with respect to the MW halo (see their Fig. 6). On the contrary, using semi-analytical models (starting with the extended Press-Schecter formalism and using various prescriptions for star formation, feedback and mixing of the stellar ejecta) Tumlinson (2006) and Salvadori, Schneider and Ferrara (2007) find rather good agreement between the model and observed halo metallicity distribution.

Independently of their success or failure in reproducing the observations, those recent models provide little or no physical insight into the physical processes that shaped the metallicity distribution of the MW halo. Indeed, if the halo was built from a  large number of sucessive mergers of sub-haloes, why is its metallicity distribution so well described by the simple model with outflow (which refers to a single system)? And what determines the peak of the metallicity distribution at [Fe/H]=$\sim$--1.6, which is (successfully) interpreted in the simple model by a single parameter (the outflow rate) ? Even in the succesfull models cited in the previous paragraph, no clear answer is provided to these questions.

In this work we present an attempt to built the halo metallicity distribution analytically, in the framework of the hierarchical merging paradigm. Preliminary results have already been presented in Prantzos (2007b). It is assumed that the building blocks were galaxies with properties similar to (but not identical with) those of the Local group dwarf galaxies that we observe today. In that way, the physics of the whole process become (hopefully) more clear. The plan of the paper is as follows: In Sec. 2 we present the simple model of the outflow and  its overall successfull description of observations; we discuss, in particular, some issues  related to its low- and high- metallicity tails. In Sec. 3 we present the properties of the sub-haloes that formed the MW halo. These are derived either from observations of the Local froup dwarf galaxies (the shape of the metallicity distribution, Sec. 3.1 and the effective yield, Sec. 3.2) or from the theory of large scale structure formation (the distribution function of baryonic sub-haloes, Sec. 3.3). The results of the model are presented in Sec. 3.4 and summarized in Sec. 4.

\section{The halo metallicity distribution  and the simple model}

The halo metallicity distribution for field stars was well established in the metallicity range -2.2 $<$ [Fe/H] $<$ -0.8 by two major surveys in the 90ies (Ryan and Norris 1991, Carney et al. 1996). An ongoing survey, known as the Hamburg/ESO survey, seeks to establish the form of the metallicity distribution at even lower metallicities. Its results should be considered as representative down to [Fe/H]$\sim$--3, and thus they superseed the survey of Ryan and Norris (1991) below [Fe/H]=--2. For [Fe/H]$>$--2.2, various biases affect the results of the Hamburg/ESO survey. Fig.1 displays a composite of the Ryan and Norris (1991) and the Hamburg/ESO surveys, adjusted to fit each other at [Fe/H]=--2.2. In this work we shall consider the main body of that distribution, and we shall only briefly discuss some issues related to its low- and high-metallicity tails.

As already noticed a long time ago (Hartwick 1976) the halo metallicity distribution is nicely described by the simple model of galactic chemical evolution (GCE). In the framework of that model, the metallicity $Z$ is given as a function of the gas fraction $\mu$ as
\begin{equation}
Z \ = \ p \ ln\left({\frac{1}{\mu}}\right) \ + Z_0 
\end{equation}
where $Z_0$ is the initial metallicity of the system (zero for all isotopes, except those created in the Big Bang) and $p$ is the {\it yield} of a stellar generation, i.e. the newly created mass of a metal returned to the ISM per unit mass blocked in stars (both long lived stars and stellar remnants). 
Metallicities and yield are expressed in units of the solar metallicity
\zs. The yield depends on the  stellar nucleosynthesis models and on the
adopted stellar initial mass function (IMF). If the system evolves at a
constant mass (closed box), the yield is called the {\it true yield},
otherwise (i.e. in case of mass loss or gain) it is called the {\it
effective yield}; its value is then always lower than the one of the
true yield and it can be analytically evaluated in some specific cases
(see e.g. Edmunds 1990).

Taking into account that the number of stars is proportional to the stellar
mass (for a constant IMF) and that the star mass fraction $\mu_*$ is
given by $ \mu_* \ = \  1. \ - \ \mu $,
one may use Eq. (1) to derive the {\it differential metallicity
distribution} (DMD) of the system as:
\begin{equation} 
{{d(n/n_1)}\over{d(logZ)}} \ = \
 {{{\rm ln10}}\over{1-{\rm exp}\left(-{{Z_1-Z_0}\over{p}}\right)}}  \
 {{Z-Z_0}\over{p}} 
\ {\rm exp}\left(-{{Z-Z_0}\over{p}}\right)
\end{equation}
where $Z_1$ is the final metallicity of the system and $n_1$ the total
number of stars (having metallicities $\le Z_1$). This function has a
maximum for $Z-Z_0=p$, allowing one to evaluate easily the effective yield $p$ if the DMD is determined observationally.

The DMD of field halo stars peaks at [Fe/H]$\sim$--1.6, suggesting an effective halo yield  $p_{Halo} \sim$1/40 \zs \ for Fe.
This has to be compared to the true yield obtained in the solar neighborhood.
The peak of the local DMD is located in the range -0.2$<$[Fe/H]$<$0.,
depending on the adopted metallicity calibration (see e.g. Haywood 2006,
Holmberg et al. 2007). Assuming that the peak is at [Fe/H]=--0.1, it
points to a local yield of $p_{Disk}$=0.8 \zs. This value is not very different from the yield  derived for the bulge of the MW,  which evolved more or less as a closed box (as suggested from its metallicity disribution, see e.g. Fulbright et al. 2006); this is considered to be the true yield (assuming the same IMF between local disk and the bulge).
Taking into account that $\sim$60-65\% of solar Fe are produced by SNIa and only $\sim$35-40\% by SNII
(e.g. Goswami and Prantzos 2000)
it turns out that the true Fe yield of SNII during the local disk evolution
was $p_{Disk}\sim$0.32 \zs \ and,
consequently, the {\it effective Fe yield} of SNII during the halo phase (where they
dominated Fe production) was $p_{Halo}\sim$0.08 $p_{Disk}$. 

The simplest interpretation of such a difference between the
effective yields in the
halo and the local disk remains still the one of Hartwick (1976), who suggested that {\it
outflow} at a rate $F \ = \ k \ \Psi$ (where $\Psi$ is the Star Formation Rate or SFR) occured
during the halo formation. In the framework of the Simple model of GCE
such outflow reduces the true yield $p_{True}$ to its effective value $p_{Halo}=p_{True}(1-R)/(1+k-R)$
(e.g. Pagel 1997) where $R$ is the Returned Mass Fraction  ($R\sim$0.32-0.36 for most of the modern IMFs, like e.g. the one of Kroupa et al. (1993) or of Kroupa 2002).
The halo data suggest then that the ratio of the outflow to the star formation rate was 
\begin{equation}
k \ = \ (1-R) \ \left({\frac{p_{True}}{p_{Halo}}} -1 \right)
\end{equation}
and, by replacing $p_{True}$ with $p_{Disk}$ one finds that $k\sim$7-8 times the star formation rate.
This high outflow rate could be interpreted either as gas expulsion after heating
by supernova explosions, or as simple flowing of gas {\it through} the system, e.g. towards the Galaxy's bulge.

\begin{figure}
\centering
\includegraphics[width=0.49\textwidth]{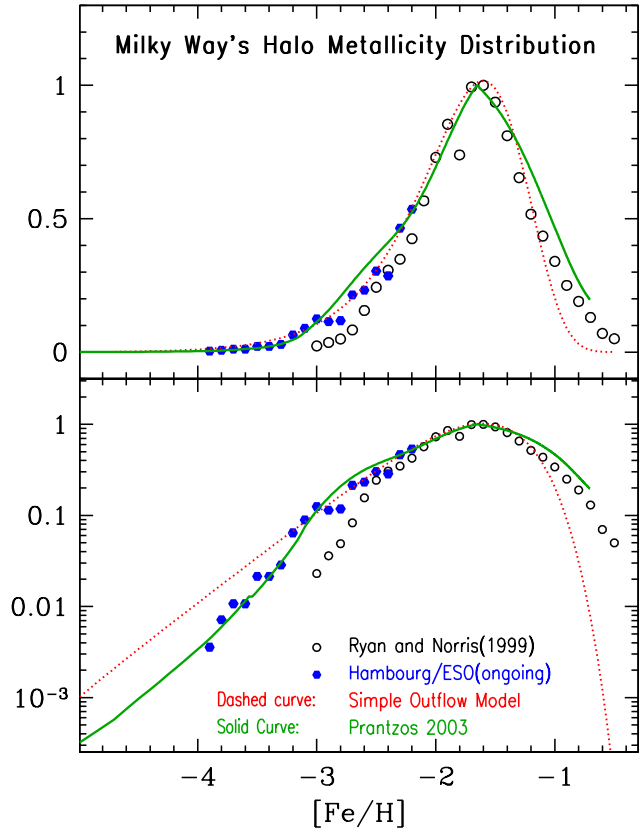}
\caption{Differential metallicity distribution of field halo stars in
 linear ({\it top}) and logarithmic ({\it bottom}) scales. Data are from
 Ryan and Norris (1991, {\it open symbols}), and the ongoing Hamburg/ESO 
 project  ( {\it filled
 symbols}). The two data sets are normalised at [Fe/H]-2.2; above that
 value, the Hamburg/ESO data are incomplete, while below that
 metallicity the Ryan and Norris (1991) data set is incomplete. 
The {\it dotted curve} is a simple model with instantaneous recycling (IRA) and  
outflow rate equal to 8 times the star formation rate. 
The solid curve is obtained as in Prantzos (2003), from a model without
IRA, an early phase of rapid infall and a constant outflow rate equal to
 7 times the SFR. All curves and data are normalised to max=1.}
\label{fig:1}       
\end{figure}

Although the siple model of GCE with outflow fits extremely well the
bulk of the halo DMD, the situation is less satisfactory for the low
metallicity tail of that DMD. The problem (a lack of low metallicity
stars w.r.t. the predictions of the simple model) was noticed  sometime ago
and prompted some interesting scenarii concerning the physics of the early galaxy, like e.g.  pre-enrichment by  a putative population of Pop. III stars (Norris 1999) or various possibilities of inhomogeneous early evolution
(Tsujimoto et al. 1999, Oey 2003).
Prantzos (2003) showed that the situation is in fact even worse if
the instantaneous recycling approximation (adopted in all previous works on the problem) is dropped and the finite
lifetimes of stars are taken into account.  Noting that the
situation is reminiscent of the G-dwarf problem in the solar
neighborhood, he further suggested that a similar solution should
naturally apply, namely an early phase of rapid infall (in a time scale
of less than 0.1 Gyr) forming the Milky Way's halo. 
Such a scenario is illustrated in Fig. 1. It may account for
the "G-dwarf" 
problem of the halo, at the price of introducing more degrees of freedom than the simple model.
At present, the situation is not quite clear regarding the low
metallicity tail of the halo DMD. We shall present some tentative
conclusions, based on preliminary data, in Sec. 3.5; however, more sound
conclusions should be drawn only after the definitive results of the
full Hamburg/ESo survey (Sch\" ork et al. in preparation).

The high metallicity tail of the halo DMD may also hold some important
clues as to the halo evolution. Its shape is surprisingly well fitted by
the simple model, up to [Fe/H]$\sim$--1. In the framework of the simple
model, this implies that the halo evolved down to a very low gas
fraction 
$\mu (\le Z_1)\sim$0.05. Indeed, since log($Z_1$)$\sim$--1 and
log($p_{Halo}$)$\sim$--1.6, the main result of the simple model (Eq. 1)
leads to $\mu \sim$0.05. However, a different interpretation was
suggested by Oey (2003). In the framework of her Simple Inhomogeneous
Model, the halo is a relatively unevolved system, composed of a
number of regions enriched to different levels (the larger ones having
lower metallicities). In that model, the final metallicity distribution
reflects essentially the size distribution of those regions, and  at
high metallicities  it scales as $Z^{-2}$, in satisfactory agreement
with observations.

At present, it is hard to distinguish between the two alternatives,
i.e. to decide whether the  strong decrease of the observed star number in the 
high metallicity tail of the halo is due to the fact that the system
is highly evolved or to its inhomogeneous nature. Obviously, it cannot
be due to observational errors alone, since the width of the
DMD ($\sim$1 dex) is much larger than the one justified by such errors(0.1-0.2 dex).  
We simply note that in the modern framework of galaxy formation through
hierarchical merging, smaller units have smaller metallicities, as
discussed in the next section.

\section{The halo DMD and hierarchical merging}

In this work we assume that the MW halo was formed by the merging of
smaller units ("sub-haloes"), as implied by the hierarchical merging scenario 
for galaxy formation.  It should be noticed at this point that, despite its
overall sucess with large scale structure in the Universe, this scenario
still faces important problems when it comes to predictions concerning the
behaviour of baryons at galactic scales.
Among those problems, the one concerning the formation chronology of 
baryonic vs 
dark matter structures (namely the fact that large galaxies are observed to
be older than small ones, while  large dark matter haloes assemble later 
than low mass ones) is known as downsizing\footnote{In fact, over the
years, the word {\it downsizing} was used to denote several different, albeit
not quite unrelated, things (see Conroy and Weschler 2008).}. The word {\it
downsizing} was first used by Cowie et al. (1996)  on observational grounds,
while the problem with spirals was 
noticed in Boissier and Prantzos (2000) and for ellipticals in Thomas et al. 
(2002). van den Bosch  (2002) showed explicitly how semi-analytical models
in the framework of hierarchical merging scenarios fail to reproduce the
colour-magnitude relation of galaxies, even if some form of feed-back is 
included. Despite several attempts, no universally acepted model meeting the
observational requirements has been proposed up to now (but see Cattaneo et 
al. 2008, for a possible solution in the case of elliptical galaxies).

Despite this important shortcoming, several models have been built in the 
framework of the hierarchical merging scenario, exploring its consequences on
the formation of the MW halo and some of them address specifically the 
issue of the halo metallicity distribution (e.g. Tumlinson 2006, Salvadori et al. 2007). In these models
star formation takes place early on in the different baryonic sub-haloes, which
subsequently  are tidally disrupted to form a halo component 
with an old stellar 
population. Such an early star formation is required to account for the
observed age of halo stars ($>$11 Gyr, e.g. Hill et al. 2002) and their
large [$\alpha$/Fe]$\sim$0.5 ratio (Cayrel et al. 2004), suggesting that those
stars had not been enriched by Fe from SNIa. 

We shall assume then that the
MW sub-haloes formed their stars early on, plausibly because at those 
early times they evolved already in a dense environment 
(i.e. the deep potential well of the MW halo); similar size sub-haloes in less
dense environments should form their stars at  a slower pace, on average, 
and develop
thus young stellar populations today. During the short period of star fomation
in each of the sub-haloes, their shallow individual potential wells could hardly
keep their gas (affected by ram-pressure stripping and heated by supernovae)
which would then leave the sub-haloes (probably to end in the bulge, as 
suggested long ago on the basis of angular momentum 
conservation arguments, e.g. Gilmore et al. 1989). Obviously, the amount of
outflowing gas (and of the metals kept inside the system, i.e. the effective
yield) should depend on the depth of the potential well of each sub-halo, 
i.e. on the mass of the corresponding dark matter sub-halo.

In order to calculate semi-analytically
the resulting DMD in the framework of the aforementionned scenario, 
it is further  assumed that each of the sub-haloes
had a DMD described well by the simple model, i.e. by Eq. (2). It
remains then to evaluate the corresponding effective yield $p(M)$ of
each sub-halo as a  function of the mass of the sub-halo, as well as 
the mass function of the sub-haloes $dN/dM$, 
in order to obtain the required result. Each one of those ingredients 
is discussed in the following sections.

\subsection{ The shape of sub-halo DMD}

It is assumed here that each one of the merging sub-haloes has a DMD
described by the simple model with an appropriate effective yield. This
assumption is based on recent observations of the dwarf spheroidal
(dSph) satellites of the Milky Way, as will be discussed below. It is
true that the dSphs that {\it we see today} cannot be the components of
the MW halo, because of  their observed abundance patterns: indeed, as
discussed in several places (e.g. Shetrone, C\^ot\'e and Sargent 2001;
Venn et al. 2004) their $\alpha$/Fe ratios are typically smaller than
the   [a/Fe]$\sim$0.4$\sim$const. ratio of halo stars. 
This implies that they evolved on longer timescales than the Galactic
halo, allowing SNIa to enrich their ISM with Fe-peak nuclei and thus 
to lower the $\alpha$/Fe ratio by a factor of $\sim$2 (as evidenced from the 
[O/Fe]$\sim$0 ratio in their highest metallicity stars, about half of the MW halo value). However, the shape of the DMD of the
simple model does not depend on the star formation history or the
evolutionary timescale. This  is precisely what makes the DMD such a
powerful diagnostic tool: it provides accurate and indepedent
information on gas flows into and out of the system, as well as on the
initial metallicity. In other terms, the question is whether the
components of the MW halo had a DMD decribed by the simple model or by
some different form. To answer that question, one may  turn either to
simulations (which are difficult to constrain) 
or to observations of galaxian systems {\it smaller} than the MW halo. This is the case of the Local group dSphs.

\begin{figure}
\centering
\includegraphics[width=0.49\textwidth]{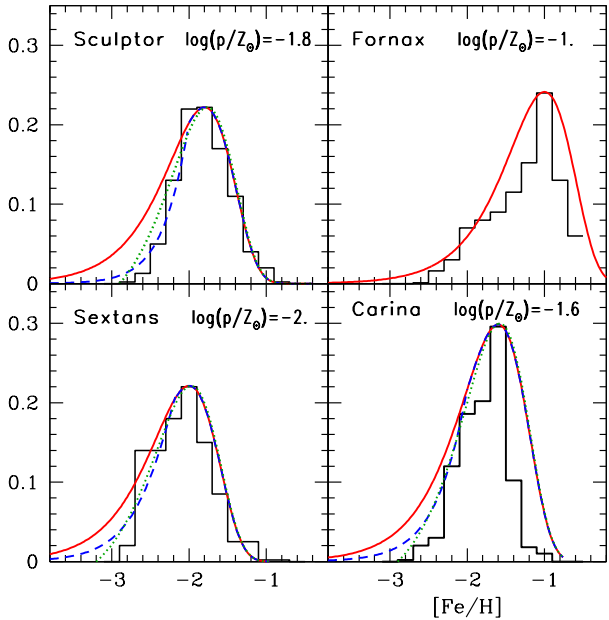}
\caption{Metallicity distributions  of dwarf satellites of the Milky Way. 
Data are in {\it histograms} (from Helmi et al. 2006). {\it Solid curves} 
indicate the results of simple GCE  models with outflow proportional to 
the star formation rate;  the corresponding effective yields (in \zs) 
appear on top right of each panel. {\it Dashed curves} are fits obtained
 with an early infall phase, while {\it dotted curves} are models with
 an initial metallicity log$(Z_0)\sim$--3; both modifications 
to the simple model
 (i.e. infall and initial metallicity) improve the fits to the data.}
\label{fig:1}       
\end{figure}

Helmi et al. (2006) report observations of several hundred red giant
stars in four nearby dSphs (Sculptor, Fornax, Sextans and Carina),
obtained with the FLAMES spectrograph in the VLT. The surface [Fe/H] ratio of those stars
should only marginally be affected by the 1st dredge-up, and thus they
may be used as tracers of the chemical evolution of the corresponding
galaxies. The resulting DMDs are displayed as histograms in Fig. 2,
where they are compared to the simple model with appropriate effective
yields ({\it solid curves}). The effective yield in each case was simply
assumed to equal the peak metallicity (Eq. 2). It can be seen that the overall shape 
of the DMDs is quite well fitted by the simple models. This is important,
since i) it strongly suggests that {\it all} DMDs of small galaxian
systems can be described by the simple model (at least, there are no
counter examples to that hypothesis) and ii) it allows to determine {\it
effective yields} by simply taking the peak metallicity of each DMD ; this will be used in Sec. 3.2 below.

Before proceeding to the determination of effective yields, we note that the fit of the simple model to the data of dSphs fails in the low metallicity tails: as already noticed in Helmi et al. (2006), there is a lack of low metallicity stars w.r.t. the simple model predictions. Helmi et al. (2006) attribute this  to a pre-enrichment of the gas out of which the dSphs were formed, i.e. they assume that $Z_0\neq$0 in Eq. (1) and (2). By fitting their data that way, they determine then initial metallicities in the range [Fe/H]$_0\sim$--2.9 - --2.7 for the four galaxies. Furthermore, they argue that this result has a strong implication ; it would imply that ``... the progenitors of the Milky way (halo) and of dSphs must have been different''. However, this is not necessarily true.

Indeed, the idea of a finite initial metallicity (pre-enrichment) was
for a long time considered as the "simplest" solution to the G-dwarf
problem in the solar neighborhood (see Pagel 1997). But, although
mathematically the simplest, it was not the most "natural", and infall
has now superseed it. In the case of the MW halo, the situation is less
clear, but early infall appears again as an interesting scenario, as
argued in Prantzos (2003) and  Sec. 3.1. In the light of these
arguments, one should not exclude that early infall is also the solution
to the "low metallicity tail" problem of the DMDs of the local
dSphs. If an early infall phase turns out indeed to be a universal
feature of galaxian systems, then the MW halo may well be composed of
galaxies similar to {\it the progenitors of the dSphs}. We shall return
to the low metallicity tail of the halo in Sec. 3.5, after considering
first the bulk of the halo DMD.

\subsection{ The effective yield of the sub-haloes}

If the DMDs of all the components of the Galactic halo are described by
the simple model, then their shape is essentially  described by the
corresponding effective yield $p$ (and, to a lesser degree, by the
corresponding initial metallicity $Z_0$). Observations suggest that the
effective yield is a monotonically increasing function of the galaxy's
stellar mass $M_*$. Data for about 40 galaxies of the Local group,
compiled by  Dekel and Woo (2003) are 
presented in Fig. 3. It can be seen that the  (median) stellar metallicity of
those galaxies varies as 0.01 $(M_*/10^6 {\rm M_{\odot}})^{0.4}$. Dekel and Woo (2003) offer a
plausible theoretical framework for that corelation, based on supernova
feedback and larger mass loss from smaller galaxies; their model
explains satisfactorily other observed properties, as e.g. velocity
dispersion and central surface brightness vs. stellar mass. This is also the relationship between the effective yield (equal to the median stellar metallicity from Eq. 2) and the mass of those galaxies. In the case of the
progenitor systems of the MW halo, however, the effective yield must have been lower, since SNIa had not time to contribute (as evidenced by the high 
$\alpha$/Fe$\sim$0.4 ratios of halo stars), by a factor of about two (see discussion in previous section). We assume
then that the effective yield of the MW halo components (accreted satellites
or sub-haloes) is given (in solar units \zs) by
\begin{equation}
p(M_*) \ = \ 0.005 \ \left({\frac{M_*}{10^6 M_{\odot}}}\right)^{0.4}
\end{equation}
i.e. the thick dotted curve in Fig. 3.

Obviously, the stellar mass $M_*$ of each of the sub-haloes should be $M_*< M_H$
where $M_H$ is the stellar mass of the MW halo. In a recent paper, Bell et al. (2007), analysing
$\sim$4 10$^6$ stars of the MW halo find that the stellar mass in the
halo region at distances 1 to 40 kpc from the Galactic center is $M_H$=
4$\pm$0.8 10$^8$ \ms \ and we adopt that value here. The
conclusions of our work  depend only slightly on the precise value of $M_H$.

An inspection of Fig. 3 shows that the effective yield of the MW halo is
substantially lower (factor $\sim$4) than the one of dwarf galaxies with stellar
masses in the range of $M_H$. Part of this difference (factor $\sim$2) is due to the fact that no SNIa contributed to the effective Fe yield of the MW halo. The remaining difference can be qualitatively understood if it
is assumed that the halo was made by the assembly of smaller mass
satellites: the resulting total mass ($M_H$) is larger than the mass of
the largest sub-halo, but the resulting mass-weighted effective yield
($p_H$) is smaller than the effective yield of the largest sub-halo. On
the contrary, if the MW halo were formed as a single system, it is hard
to understand why its effective yield is so low w.r.t. systems of
similar mass.

\begin{figure}
\centering
\includegraphics[width=0.49\textwidth]{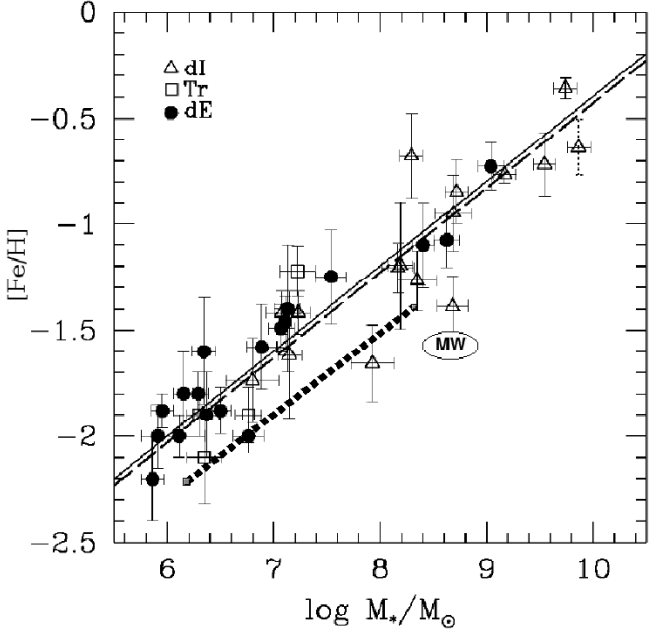}
\caption{Stellar metallicity vs stellar mass for nearby galaxies; data
 and model ({\it upper curves}) are from Dekel and Woo (2003), with {\it
 dI} standing for dwarf irregulars and  {\it dE} for dwarf ellipticals. The {\it thick 
dotted} line represents the effective yield of the sub-haloes that formed the MW halo according to this work (i.e. with no contribution from SNIa, see Sec. 3.2). The MW halo, with average metallicity [Fe/H]=--1.6 and estimated mass $\sim$4x10$^8$ \ms \ falls below both curves. }
\label{fig:3}       
\end{figure}

\subsection {The stellar mass function of the sub-haloes}

Contrary to the cases of the effective yield and of the shape of the
DMD, the mass function of the  sub-haloes which presumably formed
the MW halo can not be observed today (since
the observed number of local dwarf galaxies is insufficient for that purpose)
but has to be derived from semi-analytical or numerical calculations of structure formation.

Hierarchical galaxy formation scenarios predict  the mass function of
the dark matter sub-haloes which compose a dark matter halo at a given
redshift.
In the high resolution simulation {\it Via Lactea}, one of the most
``realistic'' simulations of a MW-like galaxy  performed so far, Diemand et
al. (2007) find that the cumulative dark sub-halo mass function at redshift $z$=0
can well be approximated by a power-law $N(>M_D)$=0.0064
($M_D/M_{DH}$)$^{-1}$ ($M_{D}$ being the mass of the dark matter
sub-haloes and $M_{DH}$ the total mass of the dark halo); this
approximation is valid over three orders of magnitude, i.e.  from 3
10$^6$ to 3 10$^9$ \ms \ inside the galactic virial radius.

The corresponding differential mass function is obviously 
\begin{equation}
\frac{dN}{dM_D} \ \propto \ M_D^{-2}
\end{equation}
Salvadori et al. (2007) find  in their numerical simulations 
that the same result is valid at higher redshifts (even up to $z$=5, see their Fig. 3), 
at least for the low mass part of the dark halo mass function.
In a recent work, Giokoli et al. (2007) find analytically (by applying the extended
Press Schecter formalism and using the progenitor (conditional) mass function) that
the result of Eq. (5) is independant of the mass of the progenitor halo  and extends down
to sub-halo masses as small as 10$^{-6}$ of that mass.

However, in our case, we are interested in the mass function of the
{\it stellar sub-haloes}, and not of the dark ones.
The initial ratio of baryonic mass $M_B$ to dark matter mass is usually taken to be
independent of the dark halo mass $M_D$ and equal to the corresponding cosmic
ratio $f=\Omega_B/\Omega_D\sim$0.17, where $\Omega_B\sim$0.04 and
$\Omega_D\sim$0.23 are the cosmic densities of baryons and dark matter,
respectively (e.g. Fukugita and Peebles 2003). If that ratio remained constant throughout galaxy
evolution, then Eq. (5) would obviously provide the required stellar
mass function also (assuming that all, or most of, baryons form stars).
However, several effects may subsequently alter that
ratio, like e.g. ram pressure stripping or galactic winds from
supernova feedback. Observations and theoretical arguments suggest that
those effects become more important in lower mass systems (e.g. Dekel and
Woo 2003). It is difficult to evaluate the impact of those effects
on the final baryonic fraction of a given galaxy, and to derive the
corresponding mass function of baryonic (mostly stellar)
sub-haloes just from first principles. We attempt here  an analytical  evaluation of
those quantities, based on the various empirical relations established
in Sec. 2 and 3.

From Eq. (3), one can see that the ratio of the outflow to the star formation rate from a galaxy (or a sub-halo) of stellar mass $M_*$ is
\begin{equation}
k(M_*) \  \propto \ p(M_*)^{-1}
\end{equation}
since $p_{True} >> p(M_*)$ for those
small galaxies/sub-haloes, which are dominated by outflows (as can be
seen in Fig. 4, top panel, even for a galaxy of $M_*\sim$10$^8$ \ms, the outflow
rate is 3 times the SFR). By virtue of Eq. (4), one has then
$k(M_*) \  \propto \ M_*^{-0.4}$, in other terms
\begin{equation}
\frac{M_{OUT}}{M_*} \  \propto \ M_*^{-0.4}
\end{equation}
where $M_{OUT}=\int F dt=\int k\Psi dt$ is the mass lost from the system and
$M_*=\int \Psi dt$ is the stellar mass of the system. On the other hand, the barionic mass $M_B$ is given by
\begin{equation}
M_{OUT} \ + \ M_* \ = \ M_{B} \ = f \ M_D
\end{equation}
with $f=\Omega_B/\Omega_D$.

For $M_{OUT}>>M_*$, Eqs (7) and (8) lead to $M_*^{0.6}\propto M_D$. Combining this with the
distribution function of the dark matter sub-haloes (Eq. 5),
one finally gets
\begin{equation}
\frac{dN}{dM_*} \ \propto \ M_*^{-1.2}
\end{equation}
i.e. {\it the distribution function of the stellar sub-haloes is flatter than
the distribution function of the dark matter sub-haloes} because of the mass loss, which becomes progressively stronger as we move to less
massive sub-haloes. The normalisation of the stellar sub-halo
distribution function is made through
\begin{equation}
\int_{M_1}^{M_2} \frac{dN}{dM_*} \ M_* \ dM_* \ = \ M_H
\end{equation}
where $M_H$=4 10$^8$ \ms \ is the adopted stellar mass of the MW halo.

\begin{figure}
\centering
\includegraphics[width=0.49\textwidth]{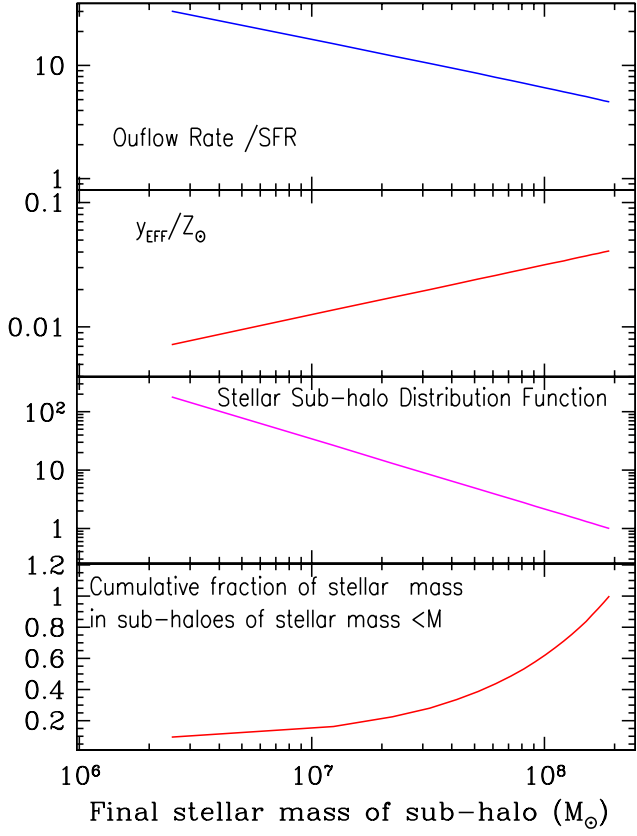}
\caption{Properties of the sub-haloes as a function of their stellar
 mass,  empirically derived as discussed
 in Sec. 3. From top to bottom: Outflow rate, in units of the
 corresponding star formation rate; Effective yield, in solar units;
 Distribution function; Cumulative fraction of stellar mass contributed
 by the sub-haloes. The total mass of the MW halo is 4 10$^8$ \ms.}
\label{fig:4}
\end{figure}

\begin{figure}
\centering
\includegraphics[width=0.49\textwidth]{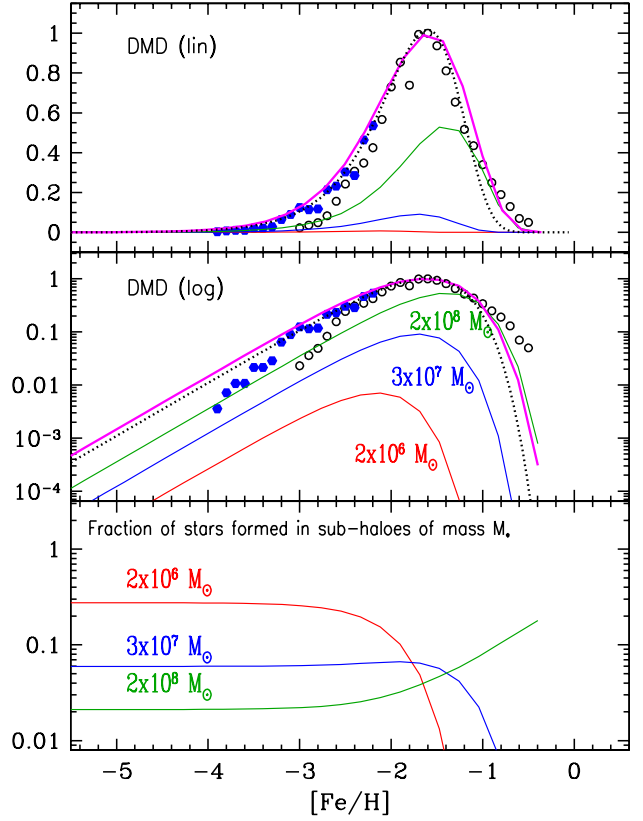}
\caption{{\it Top} and {\it middle} panels: Differential metallicity distribution (in lin and log scales, respectively) of the MW halo, assumed to be composed of a population of smaller units (sub-haloes). The individual DMDs of a few sub-haloes, from 10$^6$ \ms \ to 4 10$^7$ \ms, are indicated in the middle panel, as well as the sum over all haloes ({\it solid upper curves} in both panels, compared to observations). {\it Dotted curves} in top and middle panels indicate the results of the simple model with outflow (same as in Fig. 1). Because of their large number, small sub-haloes with low effective yields contribute the largest fraction of the lowest metallicity stars, while large haloes contribute most of the high metallicity stars ({\it bottom panel}). }
\label{fig:5}       
\end{figure}

The lower mass limit $M_1$ is adopted here to be $M_1$=1-2 10$^6$ \ms, in
agreement with the lower mass bound of observed dSphs in the Local group
(see Fig. 3). Such galaxies have internal velocities $V>$10
km/s.\footnote{Dekel and Woo (2003) calculate internal velocity as
$V=\sqrt 3 \sigma_p$, where $\sigma_p$ is the projected central velocity
dispersion.} Dekel and Woo (2003) argue that the gas in haloes with
$V<$10 km/s cannot cool to form stars at any early epoch and that dwarf
ellipticals form in haloes with 10$<V<$30 km/s. From their $V$ vs $M_*$
diagrams (their Figs. 3 and 5) one sees that a velocity of 30 km/s
corresponds to a stellar mass $M_*\sim$2 10$^8$ \ms, i.e. about half the
adopted stellar mass of the MW halo. This is the highest conceivable mass for 
a component of a $\sim$4 10$^8$ \ms halo and it is the
upper mass limit $M_2$ that we adopt here for Eq. (10). 

The main properties of the sub-halo set constructed in this section
appear in Fig. 4 as a function of the stellar sub-halo mass $M_*$.

\subsection{ The composite stellar halo DMD}

We now have all the ingredients required to calculate the DMD of the MW halo, assuming that

1) It is composed of a number of sub-haloes with stellar masses ranging from $M_1$=2 10$^6$ \ms \ to $M_2$=2 10$^8$ \ms \ and distributed as in Eq. (9); and

2) Each of  the stellar sub-haloes has a DMD given by the simple model (Eq. 2) with a mass dependent effective yield given by Eq. (4).

The resulting DMD is then obtained as a sum over all sub-haloes:
\begin{equation}
\frac{d(n/n_1)}{d(logZ)} \ = \ \int_{M_1}^{M_2} \ \frac{d[n(M_*)/n_1(M_*)]}{d(logZ)} \ \frac{dN}{dM_*} \ M_* \ dM_*
\end{equation}
The result appears in Fig. 5 (top panel in linear and middle panel in logarithmic scales, respectively). It can be seen that it fits the observed DMDs at least as well as the simple model \`a la Hartwick. In fact, at low metallicities, the behaviour of the simple model (with outflow rate = 8 times the SFR) is similar to the one of the composite model. The reason is that far from the peak (given by the effective yield) the DMDs of all individual haloes have the same slope (see middle panel of Fig. 5) and so does their sum. In the high metallicity region, the  composite model does even a little better than the simple model. This region is dominated by the DMDs of the most massive haloes, which peak at [Fe/H]$\sim$--1, whereas at low metallicities the small haloes dominate the composite DMD (see bottom panel in Fig. 5).

In summary, under the assumptions made in this section, the bulk of the
DMD of the MW halo results naturally as the sum of the DMDs of the
component sub-haloes. It should be noted that all the ingredients of the analytical model are taken from observations of
local satellite galaxies, except for the adopted mass function of the sub-haloes (which results from analytical theory of structure formation plus a small modification to account for the role of outflows). Obviously, by assuming different values for the slope of the dark matter haloes in Eq. (5) and for the mass limits
$M_1$ and $M_2$ in Eqs. (10) and (11), one may modify the peak of the resulting composite DMD, thus allowing for differences between the halo DMDs of different galaxies.

\subsection{The low metallicity tail of the DMD}

As already discussed in Sec. 2, current surveys of the low metallicity tail
of the halo DMD find a lack of stars w.r.t. the predictions of the
simple model (with outflow) below [Fe/H]$\sim$-3. A similar trend is
found in satellite galaxies of the Milky way by Helmi et al. (2006),
albeit at slightly higher metallicities ([Fe/H]$\sim$--2.8). In the case of the MW halo, Prantzos
(2003) argued that an early infall phase may account for the
discrepancy (see Fig. 1). 
Obviously, the same solution may also apply to the case of
the MW satellites, i.e. the conclusion of Helmi et al. (2006) that those
galaxies were formed from pre-enriched gas (of [Fe/H]$\sim$-2.8) does not
necessarily hold. Consequently, the main point of Helmi et al. (2006),
namely that the MW halo was not formed from sub-haloes similar to the
currently observed satellites, may not hold either. Indeed, if an early
infall phase is assumed for all the constituent sub-haloes, the
composite halo DMD will necesarily display a deficit of low metallicity
stars, in agreement with the observations. However, it is impossible at
present to determine the characteristics of the infall for each one of
the sub-haloes as a function of their mass.

For illustration purposes, we adopt here a simple toy model, where each
sub-halo undergoes an early infall phase (of primordial composition), with the smaller sub-haloes having a more important infall. This prescription for infall is motivated from the observed DMDs of the satellite dSphs of the MW: as can be seen in Fig. 2, the most massive satellites, like e.g. Fornax, are better described by the simple outflow model, whereas in the low mass satellites the simple model fails badly in the low metallicity tail and requires an important early infall. Of course, in all cases, outflow at  an appropriate rate (i.e. as to reproduce the observed effective yield) is assumed.

The results of the toy model appear in Fig. 6. Because of the assumed early infall, the low metallicity tails of the small sub-haloes are less pronounced than in the simple model. Since those sub-haloes dominate the local DMD (because of their large number and low metallicity peaks), the low metallicity tail of the composite DMD is also affected: the corresponding curve fits now the (preliminary) Hamburg/ESO survey data. Thus, it is possible that all types of  galaxian systems, namely both the components of the MW halo and the small dSphs of the local group, have undergone an early period of infall, shaping the low metallicity  tails of their DMDs.

\begin{figure}
\centering
\includegraphics[width=0.49\textwidth]{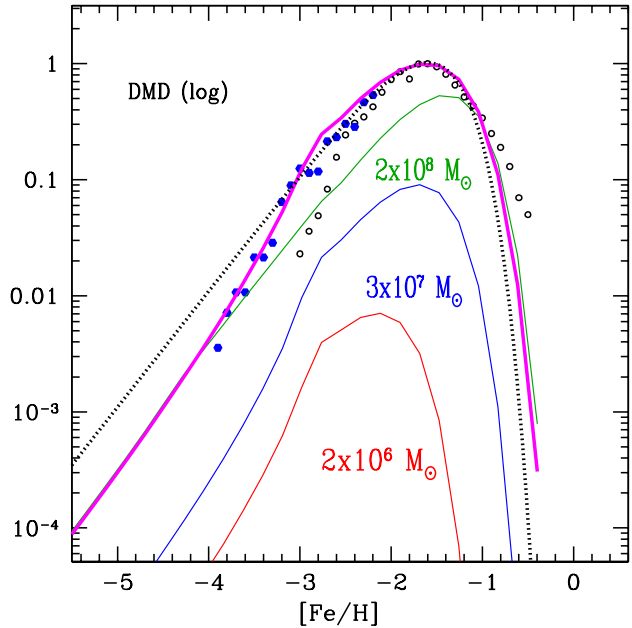}
\caption{Differential metallicity distribution of the composite MW halo population. The difference with Fig. 5 is that an early infall phase is assumed for each one of the component sub-haloes (except the most massive ones, see Sec. 3.5), as to produce a paucity of low metallicity stars. The {\it dotted curve} is the simple outflow model; data is as in Fig. 1.}
\label{fig:3}       
\end{figure}

\section{Summary}

In this work we present an analytical model for the metallicity distribution of the MW halo, in the framework of the hierarchical merging scenario for galaxy formation. It is assumed that the MW halo was composed of a number of sub-haloes with stellar masses ranging from $M_1$=2 10$^6$ \ms \ to $M_2$=2 10$^8$ \ms \ and distributed as in Eq. (9). Furthermore, each of  the stellar sub-haloes has a DMD given by the simple model (Eq. 2) with a mass dependent effective yield given by Eq. (4). Those assumptions are based on observations of  dSph satellites of the MW galaxy, while the distribution function is obtained from numerical simulations of dark matter halo formation (after explicitly accounting for  modification due to outflows).

It is found that, under those assumptions, the bulk of the MW halo DMD is reproduced in a fairly satisfactory way. The advantage of that model  over the simple outflow model (proposed by Hartwick in 1976) is that it treats the MW halo not as a single entity but as a composite one, in line with the hierarchical merging scenario for galaxy formation. Furthermore, the proposed composite model  avoids the "paradox" noticed in Sec. 3.2, namely that the effective yield of the MW halo appears to be substantially lower than the  effective yields of dwarf galaxies of comparable mass.

Observations suggest that the low metallicity tail of both the MW halo and local dSphs is defficient in stars w.r.t. the predictions of the simple outflow model. Prantzos (2003) suggested that early infall may account for that defficiency in the case of the MW halo, while Helmi et al. (2006) prefer pre-enrichment (at a level of [Fe/H]=$\sim$--2.8) for the dSphs. It is shown here, with a simple toy model based on observations of dSphs, that an early infall phase can account for the low metallicity tails of all small galaxian systems and, consequently, of the  MW halo conceived as a composite system. 

The results obtained here with semi-analytical models should be
substantiated with  detailed numerical simulations of the formation of a
stellar halo in a MW-type galaxy. Such simulations suffer at present
from several drawbacks, which make it difficult for them to provide
robust predictions. For instance, the issue of star formation vs
feedback is far from being settled. Furthermore, as shown (and stressed)
in Prantzos (2003), accounting for the finite lifetimes of stars
produces substantially different results than the commonly adopted
Instantaneous recycling approximation (IRA) at the earliest epochs (and
lowest metallicities); since almost all numerical models  adopt IRA 
(exceptions being Samland et al. 1997 or Lia et al. 2002), their predictions cannot
be considered 
as accurate, even if the issue of feedback is satisfactorily resolved.

Finally, as emphasized in Sec. 2, an in-depth discussion of the
important implications of the low metallicity tail of the DMD should
rely on the final release of the results of the Hamburg/ESO survey
(Sch\" orck et al. in preparation); the toy model presented here relies on preliminary results and should only illustrate the potential of the proposed analytical model to probe the physics of the early galaxy.

{}


\begin{thebibliography}{}

\bibitem  { } Bekki, K., Chiba, M., 2001, ApJ 558, 666

\bibitem  { } Bell, E., Zuker, D, Belokurov, V., et al. 2008, ApJ 680, 295
\bibitem  { } Boissier, S., Prantzos, N., 1999, MNRAS 307, 857

\bibitem  { } Boissier, S., Prantzos, N., 2000, MNRAS 312, 398

\bibitem  { } Carney, B., Laird, J., Latham, D., Aguilar, L., 1996, AJ 112, 668

\bibitem  { } Cattaneo,A., Dekel, A., Faber, S. M., Guiderdoni, B., 2008,   (arXiv:0801.1673) 

\bibitem  { } Cayrel, R., Depagne, E., Spite, M., et al., 2004, A\&A 416, 1117


\bibitem  { } Conroy, C., Wechsler, R., 2008 (arXiv:0805.3346)

\bibitem  { } Cowie, L., Songaila, A., Hu, E. M. \& Cohen, J., G., 1996, AJ 112, 839

\bibitem[]{} Dekel, A., Woo, J.,  2003, MNRAS 344, 1131 

\bibitem[]{} Diemand, J., Kuhlen, M., Madau, P., 2007, ApJ 667, 859 

\bibitem[]{} Edmunds, M., 1990, MNRAS 246, 678

\bibitem[]{} Eggen, O., Lynden-Bell, D., Sandage, A., 1962, ApJ 136, 748

\bibitem[]{} Font, A., Johnston, K., Bullock, J., Robertson, B., 2006, ApJ, 638, 585

\bibitem[]{} Fukugita, M., Peebles, P. J. E.,   2004, ApJ 616, 643

\bibitem[]{} Fulbright, J., McWilliam, A., Rich, M., 2006, ApJ 636, 821

\bibitem {  } Gilmore, G., Wyse, R.,  Kuijken, K., 1989, ARAA 27, 555

\bibitem[]{} Giokoli, C., Pieri, L.,  Tormen, G. 2007, MNRAS submitted (arXiv:0712.1476)

\bibitem  { } Goswami, A., Prantzos, N., 2000, A\&A 359, 151

\bibitem  { } Hartwick, F., 1976, ApJ 209, 418

\bibitem[]{} Haywood, M., 2006, MNRAS 371, 176

\bibitem[]{} Helmi, A., Irwin, M., Tolstoy, E., et al., 2006, ApJ 651, L121

\bibitem[]{} Hill, V.; Plez, B., Cayrel, R., et al., 2002, A\&A 387, 560  

\bibitem[]{} Holmberg, J., Norstr\" om, B., Andersen, J., 2007, AA 475, 519

\bibitem[]{} Kroupa, P., 2002, Science 295, 82

\bibitem[]{} Kroupa, P.,  Tout, C., Gilmore, G.,  1993, MNRAS, 262, 545

\bibitem[]{} Lia, C., Portinari, L., Carraro, G., 2002, MNRAS 330, 821



\bibitem[]{} Oey, K., 2003, MNRAS 339, 849

 \bibitem  { } Pagel B., 1997, ``Nucleosynthesis and galactic chemical 
               evolution''              (Cambidge University Press)

\bibitem[]{} Prantzos, N. 2003, A\&A 404, 211

\bibitem[]{} Prantzos, N. 2007a, in "Stellar Nucleosynthesis: 50 years
	  after B2FH", C. Charbonnel and J.P. Zahn (Eds.), EAS
	  publications Series  (arXiv:0709.0833)

\bibitem[]{} Prantzos, N. 2007b, in "CRAL-2006: Chemodynamics, from first stars
	  to local galaxies", E. Emsellem et al. (Eds.), EAS
	  publications Series  Vol. 24, p. 3 (arXiv:astro-ph/0611476)

\bibitem  { } Norris J., 1999, in ``The Galactic Halo'', 
          Eds. B. Gibson, T. Axelrod and M. Putnam     (ASP Conf. Ser. 165), p. 213

\bibitem  { } Ryan S., Norris J., 1991,  AJ 101, 1865

\bibitem[]{} Salvadori, S, Schneider, R., Ferrara, A., 2007, MNRAS 381, 647

\bibitem[]{} Scanapieco, E., Broadhurst, T., 2001, ApJ 550, L39

	
\bibitem[]{}	Samland, M., Hensler, G., Theis, Ch., 1997, ApJ 476, 544

\bibitem[]{} Shetrone, M., Cote, P., Sargent, W., 2001, Apj 548, 592

\bibitem  { } Thomas, D., Maraston, C., Bender, R., 2002, Ap\&SS 281, 371 


\bibitem  { } Tsujimoto T., Shigeyama T., Yoshii Y., 1999, ApJ Let 519, L63

\bibitem[]{} Tumlinson, J., 2006, ApJ  641, 1

\bibitem[]{} van den Bosch, F., 2002, MNRAS 332, 456


\bibitem[]{} Venn, K., Irwin, M., Shetrone, M., et al., 2004, ApJ 128, 1177




\end{thebibliography}
\end{document}